\documentclass[12pt,showpacs,preprintnumbers,amsmath,amssymb,nofootinbib,superscriptaddress]{article}
\usepackage{graphicx}
\usepackage{dcolumn}
\usepackage{bm}
\usepackage{graphics}
\usepackage{amssymb}
\usepackage{amscd}
\usepackage{afterpage}
\usepackage{float,times}
\usepackage{subfigure}
\usepackage{rotating}
\usepackage{multirow}
\usepackage{epsfig}
\usepackage{theorem}
\usepackage{moreverb}
\usepackage{euscript}
\usepackage{psfrag}

\begin{document}

{\hbox to\hsize{\hfill June 2010 }}

\bigskip \vspace{3\baselineskip}

\begin{center}
{\bf \Large 
Gravity is not an entropic force}

\bigskip

\bigskip

{\bf Archil Kobakhidze \\}

\smallskip

{ \small \it
School of Physics, The University of Melbourne, Victoria 3010, Australia \\
 archilk@unimelb.edu.au
\\}

\bigskip
 
\bigskip

{\large \bf Abstract}

\end{center}
\noindent 
{\small 
We argue that experiments with ultra-cold neutrons in the gravitational field of Earth disprove recent speculations on the entropic origin of gravitation.}

\bigskip

\bigskip

\paragraph{Introduction.} Recently, E. Verlinde presented a thought-provoking paper \cite{Verlinde:2010hp} claiming that inertia and gravitation have intrinsically thermodynamic origin. This claim is in accord with some earlier observations on the possible relation between thermodynamics and gravity \cite{Jacobson:1995ab}, \cite{Padmanabhan:2009vy}. If correct, E. Verlinde's work suggests a radically different approach to gravitational interactions. Namely, the force of gravity is associated with an entropic force caused by changes in the entropy associated with positions of material bodies. The entropy, in accord with the holographic picture, is stored on holographic screens and the space is emergent between two such screens. The gradient of Newtonian potential (gravitational force) in the emergent space measures the entropy difference between the screens. In this picture, because of macroscopic (thermodynamic) nature of gravity, there is no place for gravitons and hence no need to worry about quantization of gravity and the related problems. It seems that for macroscopic material bodies E. Verlinde's theory do indeed reproduce results of ordinary Newtonian (and perhaps also Einsteinian) gravity\footnote{The gravitational force in E. Verlinde's theory is not strictly conservative, however. It can be made almost conservative by assuming that the underline thermodynamic process is approximately reversible, i.e. it is quasi-static and the system is approximately in thermodynamic equilibrium.}. However, as we will argue in this short note, for microscopic (quantum mechanical) systems the situation is different. In particular, the results of experiments with ultra-cold neutrons in the gravitational field of Earth \cite{Nesvizhevsky:2002ef} are in disagreement with E. Verlinde's idea. 

\paragraph{Gravitational bound states of neutron.} The basic idea of the experiment \cite{Nesvizhevsky:2002ef} is as follows. Ultra-cold neutrons are allowed to fall towards a horizontal mirror under the influence of the gravitational field of Earth. In the potential well the motion of neutrons are no longer continuous in the vertical direction, but rather discrete, as predicted by quantum mechanics, that is, the neutrons jump from one height to another. The standard quantum mechanical description of the particle in a gravitational potential well (quantum bouncer), which is given by the solutions to the time-independent Schr\"odinger equation:
\begin{eqnarray}
\left[\frac{\hat p_z^2}{2m}+V(z)\right]\psi_{n}(z)=E_n\psi_n(z)~,  
\label{1}
\end{eqnarray}
where
\begin{eqnarray}
\hat p_z=-i\hbar \frac{\partial }{\partial z}~,~~{\rm and} \\
V(z)=\left\lbrace 
\begin{array}{r}
 mgz~,~~z \geq 0~~
 \\ 
 \infty~,~~z < 0~,
\end{array} \right. \nonumber
\label{2}
\end{eqnarray}
$m\approx 940$ MeV/c$^2$ is the neutron mass, $\hbar \approx 6.61\cdot 10^{-16}$ eV/s is the reduced Planck constant and $g=G_NM/R^2 \approx 9.8$ m/s$^2$ is the free fall acceleration with $G_N$ being the Newton constant and $M$ and $R$ are the mass and radius of Earth, respectively. With the boundary condition $\psi(0)=0$ (perfect mirror), equation (\ref{1}) can be solved analytically by means of the Airy functions:
\begin{equation}
\psi_n(z)=N_n {\rm Ai}\left(\frac{z}{\ell}+{\rm r}_n\right)~,~~(n=1,2,3...)~,
\label{3}
\end{equation} 
where $\ell$ is the characteristic length scale, 
\begin{equation}
\ell=\left(\frac{\hbar^2}{2m^2g}\right)^{1/3}\approx 5.9~ \mu{\rm m}~,
\label{3a}
\end{equation}
${\rm r}_n=\{-2.338,~-4.088,~-5.521,...\}$ are zero's of the Airy function, ${\rm Ai }({\rm r_n})=0$, and $N_n$ are the normalization constants defined as: $N_n=\left(\int_0^{\infty}dz |{\rm Ai}\left(\frac{z}{\ell}+{\rm r}_n\right)|^2\right)^{-1/2}$. The energy corresponding to the states (\ref{3}) are given by:
\begin{equation}
E_n=mgz_n~, ~~ z_n=-\ell \cdot {\rm r}_n~.
\label{3b}
\end{equation}

In the actual experiments \cite{Nesvizhevsky:2002ef} very slow neutrons  were guided  through a slit between a horizontal mirror and an absorber at a height $h$ above the bottom mirror, and the neutron transmission $N(h)$ was measured as a function of $h$. The recorded data show stepped increase in neutron throughput at absorber's height $\sim 15$ $\mu$m, which is consistent with the existence of a lowest energy state with  $z_1^{\rm exp}=12.2\pm 0.7_{\rm stat}\pm 1.8_{\rm syst}$ $\mu$m, in very good agreement with the theoretical prediction (\ref{3b}), $z_1\approx 13.7$ $\mu$m. This stepwise behaviour of $N(h)\sim (h-z_1)^{3/2}$ for not too large $h$ is very different from the classical prediction $N(h)\sim h^{3/2}$. The data also indicate the presence of higher energy excited states, although with the reduced accuracy.

\paragraph{Neutron states in entropic gravity.} Let us return to E.Verlinde's theory of gravitation. Although the microscopic description of the theory is not exactly specified in \cite{Verlinde:2010hp} it is sufficient for us to consider the key aspects of it. One starts with a "holographic screen" ${\cal S}$ which contains macroscopically large number of microscopic states which we denote as $|i(z)\rangle$, $i(z)=1,2,...,N({\cal E}(z),z)$. Note that the spatial coordinate $z$ is an emergent one, that is, it is a macroscopic parameter and hence only macroscopic parameters such as number of micro-states $\Omega({\cal E}(z),z)$ and the total energy ${\cal E}=Mc^2+mgz$ depend explicitly on it. The screen is described by the mixed state, 
\begin{equation}
\rho_{{\cal S}}(z)=\sum_{i(z)}p_{i(z)}|i(z)\rangle\langle i(z)|
\end{equation}
Assuming that the micro-states are equally probable, $p_{i(z)}=1/\Omega({\cal E}(z),z)$, the entropy of the holographic screen is maximal:  
\begin{equation}
S(z)=-k_{\rm B}{\rm Tr}\left[\rho(z)\log\rho(z)\right]=k_{\rm B}\log \Omega({\cal E}(z),z)
\end{equation}
That is the holographic screen has the maximal entropy that can be fitted in a volume surrounded by the screen. In this regard the holographic screen is similar to the black hole horizon. 

Now, according to \cite{Verlinde:2010hp} a particle (neutron in our case) with the Compton wavelength $\lambda=\frac{\hbar}{mc}$ has no distinct identity and is described as a fragment ${\cal N}$ of the screen, located at a position $z$, if the screen-neutron distance is $|\Delta z| < \lambda$.\footnote{In what follows we assume $\Delta z =0$ if $|\Delta z| < \lambda$, since the typical distances scales in the experiment, e.g. the "size" of a neutron state $\ell = 5.9\cdot 10^{-6}$ m, are much bigger that the Compton wavelength of neutron $\lambda = 1.3\cdot 10^{-15}$ m, and therefore one can safely neglect the finite "widht" of the screen.} If the distance is $|\Delta z| > \lambda$ then the neutron-screen system is described by the tensor product, 
\begin{equation}
\rho_{{\cal N}}(z+\Delta z)\otimes \rho_{{\cal S}/{\cal N}}(z)~,
\label{v1}
\end{equation}
where ${\cal S}/{\cal N}$ denote the "coarse grained" screen, i.e., the screen ${\cal S}$ at $z$ without the fragment ${\cal N}$. The states $\rho_{{\cal N}}(z+\Delta z)$ and $\rho_{{\cal S}/{\cal N}}(z)$ are assumed to be uncorrelated, and hence the total entropy of the neutron-screen system is a sum of the entropy associated with neutron $S_{\cal N}(z+\Delta z)$ and the entropy $S_{{\cal S}/{\cal N}}(z)$ of the "coarse-grained" screen located at $z$, $S_{\cal N}(z+\Delta z)+S_{{\cal S}/{\cal N}}(z)$. On the other hand, the state (\ref{v1}) must coincide with the state of a screen located at $z+\Delta z$, hence we have:
\begin{equation}
S_{\cal N}(z+\Delta z)=S_{{\cal S}}(z+\Delta z)-S_{{\cal S}/{\cal N}}(z)
\label{v2}
\end{equation}    
For small $\Delta z$ we can write $S_{{\cal S}}(z+\Delta z)\approx S_{{\cal S}}(z)+\Delta S_{\cal S}$, and, taking into account  that $S_{{\cal S}/{\cal N}}(z)=S_{{\cal S}}(z)-S_{{\cal N}}(z)$ due to the additivity of entropy, we obtain
\begin{equation}
S_{\cal N}(z+\Delta z)-S_{\cal N}(z)=\Delta S_{\cal S}~, 
\label{v3}
\end{equation} 
where \cite{Verlinde:2010hp} 
\begin{equation}
\Delta S_{\cal S}=2\pi k_B\frac{\Delta z}{\lambda}~.
\label{v4}
\end{equation}
The last equation is one of the basic equations in Verlinde's derivation of the classical Newtonian force \cite{Verlinde:2010hp}, which must hold true irrespective of the particular microscopic theory describing the physics of holographic screens. The equations (\ref{v3}) and (\ref{v4}) tell us that even if we start with a pure neutron state $\rho_{\cal N}(z)$ at some $z$, i.e., $S_{\cal N}(z)=0$, it "evolves" into a mixed state $\rho_{\cal N}(z+\Delta z)$ under the translation along $z$:
\begin{equation}
\rho_{{\cal N}}(z+\Delta z)\equiv U \rho_{{\cal N}}(z) U^{\dagger}~.
\label{v5}
\end{equation}
This is possible only if the translation operator $U\equiv {\rm e}^{-i\Delta z \hat P_z}$ is not an unitary operator, $UU^{\dagger}\neq \mathbf{1}$, that is, the generator of $z$-translations is not Hermitian, $\hat P_z^{\dagger}\neq \hat P_z$. To model this situation, we take: 
\begin{equation}
\hat P_z = \frac{1}{\hbar}(\hat p_z + \kappa)~,~~ \hat P_z^{\dagger} = \frac{1}{\hbar}(\hat p_z + \kappa^{*})~,
\end{equation}  
where $\kappa$ is a complex c-number, and $\hat p_z^{\dagger}=\hat p_z$ (that is, ${\rm Tr }\hat p_z^{\dagger}\rho={\rm Tr }\hat p_z \rho$). From (\ref{v5}) and (6) we deduce that, 
\begin{equation}
\hat p_z\equiv -i\hbar \frac{\partial}{\partial z}+2i{\rm Im}\kappa ~.
\end{equation}
Using Eqs (7), (12) and (13), (14), it is easy to obtain, 
\begin{equation}
{\rm Im}\kappa = - \pi m c 
\end{equation} 
As a result of the above considerations, the Hamiltonian operator, $\hat H=\frac{\hat p_z^2}{2m}+V(z)$ in (1)
is also modified. The modified Hamiltonian has real eigenvalues by the construction, that is, time evolution is unitary and we can talk about energy eigenstates as in the usual case\footnote{Note that, unitary time evolution of the system is also related with the assumption that the underlying thermodynamic process is reversible, see the footnote 1. }. However, the wave equation for energy eigenstates (1) now becomes:
\begin{equation}
\frac{\partial^2 \tilde \psi_n}{\partial z^2}-\frac{4{\rm Im}\kappa}{\hbar} \frac{\partial \tilde \psi_n}{\partial z}=
\frac{2m}{\hbar^2}\left(V(z)-E_n-\frac{2}{m}({\rm Im}\kappa)^2\right)\tilde \psi_n
\label{new}
\end{equation}
There are two extra terms which are determined by ${\rm Im}\kappa$.  The one on the right hand side of (\ref{new}) simply renormalizes the energy of neutron by an additive constant,
\begin{equation}
E_n=mgz_n - \frac{1}{2m}({\rm Im}\kappa)^2=mgz_n + 2\pi^2 mc^2~,
\label{shift}
\end{equation}
which, up to a funny factor $2\pi^2$, can be regarded as the relativistic rest energy of neutron. However, this constant shift is unobservable in the experiment \cite{Nesvizhevsky:2002ef}. More important is the impact of the second extra term in (\ref{new}). It leads to the exponential suppression of wave-functions compared to those in (\ref{3}), and, therefore, the spatial extent of states $\tilde \psi_n$ are smaller than those in (\ref{3}). Consequently, the slit is opaque for neutron states for lower values of $h$. 

Indeed, the exact solution to (\ref{new}) is
\begin{equation}
\tilde \psi_n(z)=\tilde N_n{\rm e}^{-2\pi\frac{z}{\lambda}}\psi_n(z)~,
\label{supp}
\end{equation}
where $\tilde N_n=\left(\int_0^{\infty}dz {\rm e}^{-4\pi\frac{z}{\lambda}} |{\rm Ai}\left(\frac{z}{\ell}+{\rm r}_n\right)|^2\right)^{-1/2}$. The exponential "squeezing factor" in (\ref{supp}) is enormous, because the neutron Compton wavelength $\lambda \approx 1.3\cdot 10^{-9}$ $\mu$m is much smaller than the characteristic size of the bound states $\ell \approx 5.9$ $\mu$m in (4). As a result, the probability to find a neutron state in the absorber at a height $h$ can be approximated by the step-function, 
\begin{equation}
\tilde P_n=\int_h^{\infty}dz |\tilde \psi_n|^{2}\approx 
\left\lbrace 
\begin{array}{r}
1~,~~~ h\lesssim\lambda 
\\
0~,~~~h > \lambda
\end{array}\right.
\end{equation}
Thus, if E. Verlinde's idea is correct, the neutrons would have to travel through the slit without substantial losses even if $\lambda < h<<< z_1$. This is in sharp contrast with the observations \cite{Nesvizhevsky:2002ef} according to which the slit is opaque for neutrons if $h<z_1$. Therefore, we are driven to the conclusion that gravity is not an entropic force.

\paragraph{Acknowledgements:}  I am indebted to Tony Gherghetta and Ray Volkas for useful comments on the manuscript and to Georgi Dvali for email correspondence. 
This work was supported in part by the Australian Research Council.


\end{document}